\documentclass[a4paper,12pt]{article}
\usepackage{graphicx}

\usepackage{amsmath,amssymb,graphicx}
\usepackage{cite}
\usepackage{color}
\usepackage{color}
\usepackage{ifpdf}
\ifpdf %if using pdfLaTeX in PDF mode
  \DeclareGraphicsExtensions{.pdf,.png,.jpg,.jpeg,.mps}
  \usepackage{pgf}
  \usepackage{tikz}
\else %if using LaTeX or pdfLaTeX in DVI mode
  \usepackage{graphicx}
  \DeclareGraphicsExtensions{.eps,.bmp}
  \usepackage{pgf}
  \usepackage{tikz}
  \usepackage{pstricks}
\fi
\usepackage{epic,bez123}
\usepackage{floatflt}% package for floatingfigure environment
\usepackage{wrapfig}% package for wrapfigure environment
\numberwithin{equation}{section}

\newcommand{\ii}{\mathrm{i}}

\newcommand{\dd}{\mathrm{d}}

\newcommand{\e}{\mathrm{e}}
\newcommand{\ket}[1]{\left|#1\right\rangle}
\newcommand{\bra}[1]{\left\langle #1\right|}

\newcommand{\I}{\mathbb{I}}

\newcommand{\hi}{\hat{\imath}} 
\newcommand{\hj}{\hat{\jmath}} 
\newcommand{\ti}{\tilde{\imath}}
\begin{document}

\title{Dirac lattice}%
\author{Corneliu Sochichiu\thanks{e-mail:
\texttt{sochichi@skku.edu}}\\
{\it University College \& Physics Dept.,}\\
{\it Sungkyunkwan University, Suwon 440-746, KOREA}
}%
%
%\subjclass{}%
%\keywords{}%

%\date{}%
%\dedicatory{}%
%\commby{}%
% ----------------------------------------------------------------
\maketitle
\begin{abstract}
 We study the emergence of Dirac fermionic field in the low energy description of non-relativistic dynamical models on graphs admitting continuum limit. The Dirac fermionic field appears as the effective field describing the excitations above point-like Fermi surface. Together with the Dirac fermionic field an effective space-time metric is also emerging. We analyze the conditions for such Fermi points to appear in general, paying special attention to the cases of two and three spacial dimensions. 
\end{abstract}
%\maketitle
%\tableofcontents
% ----------------------------------------------------------------
\section{Introduction}
% --------------------------------------------------------------------

It is firmly established, that we leave in a relativistic world. There was no slightest deviation from the Lorentz symmetry observed so far. The events around OPERA experiment recently confirmed the solid status of this fact \cite{Adam:2011zb}. This means, that any reasonable physical theory, at high enough speeds, should be relativistic, must respect Lorentz symmetry, or more generally, the Poincar\'{e} symmetry. But Poincar\'{e} or Lorentz symmetry is not a tame subject, either group is non-compact, which is a source of multiple problems in quantum theory.  

These problems  lead some authors to question the fundamental role of the relativistic field models, as well as, the exact nature of the Lorentz symmetry \cite{Jizba:2012zz}. There was a number of proposals for  models with no exact Lorentz symmetry, but which are supposed to flow to a relativistic system in the low energy limit \cite{Horava2009,Horava2009a}. The general idea is that the observable relativistic high energy behavior is only an approximation to a more complicated `super high energy' system with no Lorentz symmetry,  with remarkably small corrections at the present day available energies.

The Lorentz symmetry is known to naturally appear in the low energy description of some Fermi systems. Thus, Tomonaga--Luttinger liquid in one dimension \cite{Tomonaga:1950zz,Luttinger:1963zz}, has an effective description in terms of an interacting \( 1+1 \) dimensional fermion. In fact, Dirac fermion generically emerges in the case one-dimensional lattices. In two spacial dimensions, for long time it has been known, that the low energy description of the electronic wave function of graphene, one atom thick layer of graphite, is given by a Dirac fermion in \( 2+1 \) dimensions \cite{Semenoff:1984PhRvL} (see \cite{2009RvMP...81..109C} for a review of properties of graphene). The general conditions of emergence of Dirac fermions from two dimensional lattices were extensively studied in \cite{PhysRevB.83.245125}. The low energy theory of superfluid ${}^3$He is also known to be related to Dirac fermion \cite{volovik2009universe}. Some three dimensional substances exhibit (pseudo)relativistic Fermion behavior as well \cite{Frank2008410,PhysRevLett.106.056401}.  Thus, in \cite{Manes:2011jk} the emergence of Weyl particle dispersion relations was studied in connection with three-dimensional crystallographic groups.

In this work we address the problem of finding such microscopic systems, which in the low energy limit flow effectively to the Dirac fermion system. The starting point is a non-relativistic system of particles hopping on a graph. These particles are obeying a Pauli exclusion principle. The exclusion principle is essential here, since it is the dynamics near the Fermi point, which brings us to the Dirac particle in the low energy limit. We present here a simple argument, showing that the spectrum of fluctuations around a non-degenerate linear Fermi point is \emph{necessarily} described by Dirac particles. The only assumptions we make, are the analyticity of the energy spectrum near the Fermi point and irreducibility in the continuum limit. The last assumption is equivalent to emergence of continuous rotational symmetry. The fact, that the spectrum of fluctuations near a Fermi point is described by a (pseudo)relativistic fermion, was claimed earlier \cite{volovik2009universe}, based on the completeness of the ABS construction \cite{Atiyah19643}, but a direct proof was not known to us. On the other hand our proof presented here, is direct and rather simple. As a consequence, our task becomes to look for a system with a non-degenerate spectrum around Fermi points. We can express this as a condition imposed on the \emph{adjacency matrix} describing the graph, in addition to the condition that the graph admits a continuum description in the low energy limit. As a result, we present a general solution in two and three spacial dimensions, as well as a solution in arbitrary dimension, which we believe is general, although we could not find the proof of that fact in this work.

The plan of the remainder of this paper is as follows. In the next section, we set up the problem and introduce the notations. Then, in section \ref{sec:FermiPts}, we show that non-degenerate fluctuations around a Fermi point in the linear order in momentum are described by a Dirac fermion or a collection of Dirac fermions. In the last case, the energy spectrum of fluctuations belongs to a product of inequivalent Clifford algebra representations. In this case, the rotational symmetry in the continuum limit is generally broken. Restricting to a single copy of such representations, we restore the rotational symmetry. There we also show, that an effective (Minkowski) geometry emerges in this limit. The conditions, that we stay on a Fermi point, are formulated in the form of algebraic equations \eqref{cond-coord}, whose solution is considered in section \ref{sec:Gammas}. There we give a general solution for the expansion coefficients in the cases of graphs flowing to two and three dimensional spaces. We present also a solution for the general case of arbitrary dimension. Finally, in section \ref{sec:conclusion}, we discuss the obtained results, point to perspectives and generalizations as well as further problems. In the appendix, we construct the the holomorphic Clifford algebra basis, which is used to expand the adjacency matrix from the main part.

% --------------------------------------------------------------------
\section{The model}\label{sec:gen}

Let us consider a the model of a fermionic particle hopping on a graph. The energy of hopping between two sites, say \( x \) and \( y \) is given by the element  \( t_{<xy>} \) of the adjacency matrix  \( \mathbf{T} \). Further properties of the graph in the form of  restrictions imposed on the adjacency matrix we will specify later. So far, the most general tight-binding Hamiltonian describing this system is given by,
\begin{equation}\label{hamiltonian1}
 	H=\mathbf{a}^\dag\cdot \mathbf{T}\cdot \mathbf{a} \equiv \sum_{<xy>} t_{<xy>} a^{\dag}_x a_{y},
\end{equation}
where \( a^{\dag}_x \) and \( a_x \) are, respectively creation and annihilation operators of a fermionic particle at the site \( x \). In what follows we assume that the Fermi sea is filling half of the available energy states of this model.

The hermiticity of the Hamiltonian requires that the the amplitudes satisfy $t_{<xy>}=t_{<yx>}^*$ where the star denotes the complex conjugation. On the other hand, the complex phase in the transition amplitude \( t_{<xy>} \) can be related through Peierls substitution to a background (electro)magnetic field, therefore in the absence of such a field it is natural to restrict the model to real valued transition amplitudes, \( t_{<xy>}=t_{<xy>}^*=t_{<yx>} \).

In order to admit a continuum description in the low energy limit, the graph should possess at least an approximate translational symmetry. As a first approach, here we consider graphs possessing the periodicity of a \( D \)-dimensional lattice, leading to the flat space continuum limit. The periodicity property of the graph is reflected in the following block structure of the adjacency matrix \( \mathbf{T} \): There is a primitive cell subgraph which consists of \( p\) vertices, which are periodically reproduced in the form of a \( D \)-dimensional lattice with lattice vectors \( \hi  \). The primitive cell itself is an abstract graph, not necessary a part of a physical lattice. The adjacency matrix block corresponding to primitive cell is a \( p \times p \) matrix \( M \). In addition there are transitions between neighbor primitive cells \( \mathbf{n} \) and \( \mathbf{n}+\hi \), for  \( i= 1,\dots, D \).  These transitions are given by \( D \) blocks \( \Gamma_i \) also of size \( p\times p \). 

In this work we mostly keep the primitive cell index as an implicit matrix index, while the lattice coordinate \( \mathbf{n} \)  will be shown explicitly.

Taking all above into account, we can rewrite the Hamiltonian \eqref{hamiltonian1}, using explicitly the block structure of the adjacency matrix,
\begin{equation}\label{transl-inv}
 	H=\sum_{\mathbf{n},i}\left(
	a^\dag_{\mathbf{n}+\hi }\Gamma_i a_{\mathbf{n}}+
	a^\dag_{\mathbf{n} }\Gamma_i^\dag a_{\mathbf{n}+\hi}
	\right)+
	\sum_{\mathbf{n}}a^\dag_{\mathbf{n}} M a_{\mathbf{n}}.
\end{equation} 

The Hamiltonian in the form of \eqref{transl-inv} contains a discrete translational symmetry and, therefore,  is suitable for Fourier transform,
\begin{equation}
 	a( \mathbf{K})=\sum_{\mathbf{n}}a_{\mathbf{n}}\e^{\ii \mathbf{K}\cdot \mathbf{n}},\qquad 
	a_{\mathbf{n}}=\int_{\mathcal{B}}\frac{\dd^{D} {K}}{(2\pi)^D} a(\mathbf{K})\e^{-\ii \mathbf{K}\cdot \mathbf{n}},
\end{equation}
where the momentum $k$ is integrated over the first Brillouin zone $\mathcal{B}$, defined by
\begin{equation}
 	\mathbf{K}=K_i \ti,\quad -\pi \leq K_i <\pi,
\end{equation}
where the tilded vectors $\ti$, $i=1,\dots, D$ form the dual  lattice  basis,
\begin{equation}
 	\ti\cdot \hj=\delta_{ij}. 
\end{equation}

In terms of Fourier transform the Hamiltonian takes the form,
\begin{equation}\label{ham-micro-four}
 	H=\int_{\mathcal{B}}\frac{\dd^{D} {K}}{(2\pi)^D} \,
	a^\dag(\mathbf{K})\left(\sum_{i}
	\left(\Gamma_i\e^{\ii K_i}+\Gamma_i^\dag\e^{-\ii K_i}\right)
	+M
	\right)
	a(\mathbf{K}),
\end{equation}
where the Fourier modes $a(\mathbf{K})$ and $a^\dag(\mathbf{K})$ are a $p$-dimensional vector and a co-vector, respectively. The quantity in the parentheses,
\begin{equation}\label{energy-matrix}
 	 h(\mathbf{K})=\left[\sum_{i}
	\left(\Gamma_i\e^{\ii K_i}+\Gamma_i^\dag\e^{-\ii K_i}\right)
	+M\right],
\end{equation}
is called \emph{energy matrix}. The eigenvalues of this matrix give the energy modes of the Hamiltonian. The properties of \( h(\mathbf{K}) \) is the main subject of the next section.

% --------------------------------------------------------------------
\section{Low Energy Limit: Fermi points}\label{sec:FermiPts}

The low energy behavior of the model is determined by the regions of the momentum space near the Fermi surface. Therefore, the geometry of the Fermi surface is essential for the properties of the model this limit. In a generic situation, the Fermi surface is a surface of co-dimension one, i.e. a \( (D-1) \)-dimensional surface in the \( D \)-dimensional  space. However, the Hamiltonian  may contain terms, leading to degeneracy of the Fermi surface down to a lower dimensional surface or a gapped state. 

The Atyah--Bott--Sapiro construction \cite{Atiyah19643}, or just symmetry arguments, imply that the emergence of Dirac fermion in the low energy limit is conditioned by the presence of a Fermi point \cite{volovik2009universe}. Therefore, let us concentrate on finding the conditions of appearance of Fermi points and Dirac fermions.

Most generally, a  Fermi point \( \mathbf{K}^* \) is defined by two conditions,
\begin{equation}\label{FermiPoint}
 	\det h(\mathbf{K}^*)=0,\qquad \det[h(\mathbf{K}^*+ \mathbf{k})]\neq 0,\text{ for }\mathbf{k}\neq 0.
\end{equation}
The first equation means that there are zero modes at \( \mathbf{K}^{*} \), while the second one implies that this is an isolated point. More specifically, we consider linearly non-degenerate Fermi points, which imply that for given choice of coordinates in the momentum space, the energy matrix is non-degenerate in the linear approximation, i.e.
\begin{equation}
 	\det[h(\mathbf{K}^{*})+ \alpha_{i}(\mathbf{K}^{*})k_{i}]\neq 0,\text{ for } \mathbf{k}\neq 0,	
\end{equation}
where \( \alpha_{i}(\mathbf{K}^{*})= \partial_{i} h(\mathbf{K}^{*}) \).

In principle, the energy matrix \( h(\mathbf{K}) \) can be diagonalized, which gives the energy bands \( \varepsilon_n(\mathbf{K}) \), where \( n=1,\dots, p \). The Fermi point condition \eqref{FermiPoint} implies that some of these bands must cross zero at the Fermi point \( \mathbf{K}^* \), while other bands may have a gap. In the low energy limit the states belonging to the bands having gap will be frozen. Therefore, only gapless bands are relevant for this limit. So, one can locally project from the total \( p \)-dimensional ``internal space'' to the subspace of gapless modes only. The resulting geometry will be generically a fiber bundle, or a K-theory element, as described in  \cite{PhysRevLett.95.016405}. Since the contribution of gapped bands is trivial in the low energy limit, in this work we assume a minimal case with no gapped bands.

If we expand the energy matrix in the vicinity of the Fermi point,
\begin{equation}
 	 h(\mathbf{K}^*+ \mathbf{k})= \alpha_i(\mathbf{K}^*) k_i +O({k}^2),
\end{equation}
where the matrices \( \alpha_i(\mathbf{K}^*) \) are given by,
\begin{equation}\label{alpha-matr}
 \alpha_i(\mathbf{K}^*)=
	\ii \left(\Gamma_i\e^{\ii K_i^*}-\Gamma_i^\dag\e^{-\ii K_i^*}\right).
\end{equation}
then for small enough nonzero \( k_i \), the condition \eqref{FermiPoint} becomes,
\begin{equation}\label{FermiPoint2}
 	h(\mathbf{K}^*)=0, \qquad \det[\alpha^i(\mathbf{K}^*) k_i]\neq 0.
\end{equation}
One might suspect, that matrices \( \alpha_{i} \), satisfying the non-degeneracy condition \eqref{FermiPoint2} should be related to Dirac matrices. Inverse is definitely true. In what follows, let us study the properties of these matrices and show that it is indeed the case.

% --------------------------------------------------------------------
\subsection{The properties of matrices \( \alpha_i \)}
% --------------------------------------------------------------------
The problem of finding a non-degenerate set of \( D \) matrices \( \alpha_i \), \( i=1,\dots,D \) can be related to the geometrical problem of finding the maximal number of linearly independent vector fields on a sphere \cite{Adams:1962}.   The solution can be expressed in terms of Clifford algebras by Atiyah--Bott--Shapiro (ABS) construction \cite{Atiyah19643} (see also \cite{SpinGeometry}). 

Still there is a simple way to see, that the second condition \eqref{FermiPoint2} implies that the matrices \( \alpha_i \) are generators of Clifford algebra. Consider the eigenvalue problem for \( \alpha^ik_i \). It is not difficult to see, that in the vicinity of Fermi point  the leading term in \( k \) should take take the form,
\begin{equation}
 	\varepsilon_{n}(k)=\pm | \xi_n \cdot \mathbf{k}|+O(k^2),
\end{equation}
where \( \xi_n \) are some non-degenerate \( D \)-dimensional square matrices. The absolute value denotes usual Euclidean norm,
\[
 	|y|=\sqrt{ \mathbf{y}^2}.
\]

Restricting to the subspace corresponding to one \( \xi_n \), we have,
\begin{equation}
 	\varepsilon^2_n \approx (\xi_n \cdot k)^2+\dots,
\end{equation}
which means that the leading term of the restriction of the energy matrix to the subspace of fixed \( \xi_n \) must be proportional to the identity matrix (in cell index),
\begin{equation}\label{Cliff-alpha}
 	\left(\alpha_{n}^i(K^*) k_i\right)^2=(\xi_n \cdot k)^2=g_n^{ij}k_ik_j,
\end{equation}
where we introduced the \( n \)-th subspace ``metric tensor'' given by the matrix \( g_n=\xi_n^T \cdot \xi_n \) and \( \alpha_{n}^{i} \) is the restriction\footnote{It is a simple exercise to show that the \( n \)-subspace is an invariant space for each \( \alpha^{i} \), therefore such a restriction is well-defined.}  of matrix \( \alpha^{i} \) to the \( n \)-subspace. Furthermore, the equation \eqref{Cliff-alpha}  implies that the matrices \( \alpha^i \), when restricted to \( n \)-subspace, should obey Clifford algebra anti-commutation relations,
\begin{equation}\label{alpha-Cliff}
 	\alpha_{n}^i \alpha_{n}^j+\alpha_{n}^j \alpha_{n}^i=2 g_n^{ij},
\end{equation}
where \( g_{n}^{ij} \) is a positive definite matrix in \( ij \) indices.
This means that in a general basis matrices \( \alpha_i \) satisfy the anti-commutation relations, 
\begin{equation}
 	\{\alpha^i,\alpha^j\}=2 G^{ij}
\end{equation}
with diagonalizable  \( p \times p \) matrices \( G^{ij}  \), \( i,j=1,\dots, D \) with eigenvalues \( g_n^{ij} \).

In the case of distinct \( g_n \)'s, the \( D \)-dimensional rotational symmetry remains broken even in the continuum limit unless all \( g_n^{ij} \) are equivalent up to a scale factor, \( g_n^{ij}=g^{ij}t_n^2 \). The scale factor as well as the metric \( g_{ij} \) should be positive in order to keep the positivity of the energy-square. This results in the structure of \( G^{ij} \),
\[	
 	G^{ij}=g^{ij}T^2,
\]
where \( T \) is a \( N \times N \) matrix with the eigenvalues \( t_n \) in the diagonal basis. This implies that in some bases we can split the matrices \( \alpha_i \) according to,
\begin{equation}
 	\alpha_i=\tilde{\alpha}_i \otimes T,
\end{equation}
where \( \tilde{\alpha}_i \) realize an irreducible representation of the \( D \)-dimensional Clifford algebra,
\begin{equation}\label{Cliff-genuine}
 	\{\tilde{\alpha}^i,\tilde{\alpha}^j\}=2 g_{ij}.
\end{equation}

Now let us recall, that in the absence of gapped modes, the dimension of  \( \alpha_i \)-matrices is given by  \( p \), the number of elements in the primitive cell. Since the structure of \( \alpha_i \)-matrices is a product of a Clifford algebra-valued matrix to an arbitrary non-degenerate hermitian matrix \( T \), we can evaluate the required number of elements in a primitive cell which can yield a relativistic fermion in the low energy limit, using the dimensions of respective Clifford algebra representations \cite{SpinGeometry}. 

Thus, for \( D \)-dimensional Weyl fermion one has,
\[
	p_{\text{Weyl}}=N 2^{[D/2]-1},
\]
for some positive integer number \( N \). Dirac representation implies,
\[
 	p_{\text{Dirac}}=N2^{[D/2]}.
\]
Minimal case would be when \( N=1 \). As we will see below, the Fermi point in our model is always Dirac point, unless additional constraints are imposed. Therefore, the necessary condition for a graph to be Dirac lattice is to contain at least \( 2^{[D/2]} \) elements in the primitive cell. There are not many Bravais lattices satisfying this criteria, but detailed analysis of possibilities goes beyond the scope of this work.

% --------------------------------------------------------------------
\subsection{Clifford algebra basis and the low energy limit}
% --------------------------------------------------------------------
As we just established, the Fermi point condition implies that the matrix-valued coefficients \( \alpha_i(\mathbf{K}^*) \) should be generators of a \( D \)-dimensional Clifford algebra or a direct sum of \( D \)-dimensional Clifford algebras. Furthermore, the SO(\( D \)) symmetry in the low energy limit requires that it should be copies of the same Clifford algebra.

For the further analysis of the structure of matrices \( \alpha_{i}(\mathbf{K}^{*}) \) it is convenient to expand them in terms of a standard complex basis of a \( D' \)-dimensional Clifford algebra, consisting of matrices \( \Sigma_I \), \( I=1,\dots D'/2 \),  satisfying the following anti-commutation relations, 
\begin{equation}\label{sigma-matr}
 	\{\Sigma_I,\Sigma_J\}=\{\Sigma_I^\dag,\Sigma_J^\dag\}=0,\qquad
	\{\Sigma_I,\Sigma_J^\dag\}=\delta_{IJ}\I.
\end{equation}
In addition, each \( \Sigma_I \) is a matrix with real entries. It is clear that the even dimension \( D' \) of the Clifford algebra \eqref{sigma-matr} should be equal or greater than the lattice dimension \( D \). Therefore, the minimal value of \( D' \) is given by, \( D'_{\text{min}}=2\lceil D/2 \rceil\), where \( \lceil \cdot \rceil \) denotes the ceiling function. The complete basis of matrices satisfying these conditions is constructed in the Appendix \ref{app:Cliff}.

It is not difficult to convince oneself that the matrices \( \Gamma_i \) too, can be expanded in terms of basic matrices \( \Sigma_I \) with real coefficients, i.e., \footnote{We use convention, according to which summation is assumed over the repeated index \( I \).}
\begin{equation}
 	\Gamma_i=\Gamma_{iI}\Sigma_I,\qquad \Gamma_i^\dag=\Gamma_{iI}\Sigma_I^\dag.
\end{equation}

Next we can observe, that the consistency of the Fermi point condition \eqref{FermiPoint2} implies that the intracell matrix \( M \) should be expandable in terms of hermitian parts of \( \Sigma_I \),
\begin{equation}
 	M=m_I(\Sigma_I+\Sigma_I^\dag),
\end{equation}
where \( m_I \) are real.

In terms of the $\Sigma_I$-basis the Fermi point equation takes the form,
\begin{equation}\label{cond-coord}
 	h_I( \mathbf{K})=0,\qquad I=1,\dots, D'/2,
\end{equation}
where,
\begin{equation}
 	 h_I(\mathbf{K})\equiv\sum_{i}\Gamma_{iI}\e^{\ii K_i}+m_I.
\end{equation}

The coefficients \( \Gamma_{iI} \) should be regarded as parameters of equations \eqref{cond-coord}, allowing only point-like solutions for the momentum  \( \mathbf{K} \). It is clear, that if $ \mathbf{K}^*$ is a solution to the Fermi point condition \eqref{cond-coord}, then, due to the symmetry with respect to complex conjugation, $- \mathbf{K}^*$ is a solution too. These two  solutions carry opposite topological charges \cite{VolovikLect.NotesPhys.718:31-732007}, such that the net charge for the pair vanishes. Therefore, such a pair represents a minimal set of Fermi points. As a degenerate case one can have a single point \( \mathbf{K}^{*}=0 \) with zero charge, but such a configuration may suffer from topological instability. 

We postpone the further discussion on the problem of finding appropriate $\Gamma_{iI}$ and \( m_I \),  till the Section \ref{sec:Gammas}. So far let's assume that the equation \eqref{cond-coord} admits a minimal solution with a single pair of Fermi points, which we will denote $\pm \mathbf{K}^*$. 

In the vicinity of either Fermi point \(\pm \mathbf{K}^* \) the matrices \(  \alpha_{i} \) take the form,
\begin{equation}\label{alpha-1}
 	\alpha_{ i}(\pm \mathbf{K}^*)=\ii \Gamma_{iI}
	\left(\cos K_{ i}^* (\Sigma_I-\Sigma_I^\dag)
	\pm \sin K_{ i}^* \ii (\Sigma_I+\Sigma_I^\dag)
	\right),
\end{equation}
where \( \mathbf{k} \) is the offset momentum in the vicinity of  \( \pm \mathbf{K}^{*} \), i.e. \( \mathbf{K}=\pm \mathbf{K}^{*}+ \mathbf{k} \). The energy matrix in these vicinities takes the form,
\begin{equation}
 	h_{\pm}(\mathbf{k})\equiv h(\pm \mathbf{K}^{*}+ \mathbf{k})= \alpha_{i}(\pm \mathbf{K}^{*})k_{i}+O(k^{2}).
\end{equation}

Introducing the subspace associated with the sign $\pm$ of the Fermi point, we can re-write the matrices \eqref{alpha-1} in the following form,
\begin{equation}\label{alpha-ht}
 	\hat{\alpha}_{i}=
 	\Gamma_{iI}
	\left(\cos K_{i}^* \ii (\Sigma_I-\Sigma_I^\dag)\otimes \I
	- \sin K_{i}^*  (\Sigma_I+\Sigma_I^\dag)\otimes \sigma_3
	\right).
\end{equation}

Thus,  matrices $\hat{\alpha}_{i}$ are linear combinations of Hermitian matrices $\beta_a$, $a=1,\dots, D'$, defined as,
\begin{equation}\label{cliff-alg-matr}
 	\beta_{2 I-1}=-(\Sigma_I+\Sigma_I^\dag)\otimes \sigma_3,
	\quad
	\beta_{2 I}=\ii (\Sigma_I-\Sigma_I^\dag)\otimes \I,
	\qquad I=1,\dots, D'/2,
\end{equation}
i.e.,
\begin{equation}
 	\hat{\alpha}_{ i}=\xi_{ i}^a \beta_a,
\end{equation}
where,
\begin{equation}\label{embedd}
 	\xi_{i}^{2I-1}=\Gamma_{iI}\sin K^{*}_{i}, \quad
	\xi_{i}^{2I}=\Gamma_{iI}\cos  K^{*}_{i}.
\end{equation}

The matrices $\beta_a$ realize a standard $D'$-dimensional Clifford algebra basis,
\begin{equation}\label{cliff-alg}
	\{\beta_a,\beta_b\}=2\delta_{ab}\I.
\end{equation}
Therefore the coefficients $\xi_{i}^a$ can be regarded as vielbein coefficients for the embedding of a $D$-dimensional  plane into $\mathbb{R}^{D'}$, where the Clifford algebra \eqref{cliff-alg} is defined. 
This embedding produces an induced metric on the \( D \)-dimensional plane, given in terms vielbeins \( \xi_i^a \),
\begin{equation}\label{ind-metric}
 	g_{ij}=\xi_i^a \xi_j^a.
\end{equation}

The embedding coefficients and, therefore, the induced metric \eqref{ind-metric} are inherited from the adjacency matrix elements \( t_{<xy>} \) in the Hamiltonian \eqref{hamiltonian1}. Therefore, even in the case when the graph is a lattice embedded in the Euclidean space, the `induced metric' \eqref{ind-metric} does not necessarily coincide with the real space metric. They coincide, however, if the transition amplitudes on the graph are inversely proportional to the real space distance between the nodes, at least in the leading order. This is often the case for the lattices of physical substances, consisting of single chemical component, like e.g. graphene, where this condition is satisfied and the induced geometry is indeed equivalent to the real space geometry, up to a scaling factor given by the Fermi speed. In more general case, the induced metric \eqref{ind-metric} can be interpreted  as an anisotropic or tensor valued  `Fermi velocity'.

To obtain the standard Dirac fermion action, let us take the following steps. First, introduce an `orthogonal basis' with respect to the induced metric \eqref{ind-metric}. If we choose the Cartesian basis of the embedding space such that basis vectors with numbers \( a'=1,\dots,D \) span our \( D \)-dimensional space, while \( a_{\bot}=D+1,\dots, D' \) belong to the orthogonal completion, then the Cartesian momentum in this basis takes the form,
\begin{equation}
 	q^{a'}=\xi_i^{a'} k_i, \qquad a'=1,\dots,D.
\end{equation}
  
In terms of the Cartesian momentum the low energy Hamiltonian \eqref{ham-micro-four} takes the form,
\begin{equation}\label{low-energy-ham}
	 H=J\int \frac{\dd^D q}{(2\pi)^D}\, \psi^\dag(q) \beta_{a'} q^{a'} \psi(q),
\end{equation}
where \( J\) is the square root of the determinant of the induced metric
\begin{equation}
 	J=\sqrt{\det \|g_{ij}\|}, 
\end{equation}
and $\psi(q)$ and $\psi^\dag(q)$ include only the low energy modes of, respectively, $a$ and $a^\dag$. Matrices $\beta^{a'}$ belong to the reduction of the \( D' \)-dimensional Clifford algebra \eqref{cliff-alg} to the $D$-dimensional subspace. 

Taking into account the commutation relations, we can write down the corresponding low energy effective action,
\begin{equation}\label{low-energy-act}
 	S_{ \mathrm{le}}=
	J\int \frac{\dd t\dd^D q}{(2\pi)^D}\,
	\ii \psi^\dag(q) \dot{\psi}(q)-\int\dd t\,H(\psi^\dag,\psi).
\end{equation}

In the low energy theory one can extend the momentum integration to entire momentum space $\mathbb{R}^D$. Making the inverse Fourier transform, we arrive to real space low energy action,
\begin{equation}
 	S_{\mathrm{le}}=\int\dd^{D+1}x\, 
	\left(
	\ii\psi^{\dag}\dot{\psi}-\ii\psi^\dag \beta^{a'} \partial_{a'}\psi
	\right),
\end{equation}
where $\partial_{a'}=\partial/\partial x^{a'}$ is the partial derivative with the respect to the induced `Cartesian' coordinate. Also, in order to absorb the Jacobian factor \( J \), we rescaled the field \( \psi\to J^{-1/2}\psi \).

As the next step, let us generate the Dirac matrix \( \gamma^{0} \) from matrices $\beta^a$ given by \eqref{cliff-alg-matr}, as follows,
\begin{equation}\label{gamma0}
 	\gamma^0= \ii^{D'/2}\beta_1 \beta_2\cdots \beta_{D'}.
\end{equation}
By the construction this matrix anticommutes with all $\beta^a$, as well as satisfies,
\begin{equation}
 	(\gamma^0)^2=-1.
\end{equation}
Then, define the Dirac conjugate,
\begin{equation}
 	\bar{\psi}=\psi^\dag \gamma^0,
\end{equation}
as well as the spacial Dirac matrices \( \gamma^{a'} \),
\begin{equation}
 	\gamma^{a'}=\gamma^0 \beta^{a'}.
\end{equation}

Once this is done, the low energy effective action takes the `genuine' relativistic Dirac form,
\begin{equation}\label{genuine-Dirac}
 	S_{ \mathrm{le}}=
	-\ii\int\dd^{D+1}x\, \bar{\psi}\gamma^\mu \partial_\mu \psi.
\end{equation}

Let us note, that the low energy effective action \eqref{genuine-Dirac} is written in the orthonormal frame with respect to the \emph{induced metric} \eqref{ind-metric}. The Dirac matrices in this frame satisfy canonical anti-commutation relations. In the case in which the graph is a lattice embedded in real space we can go to the real space Cartesian basis using usual basis transformation rules. In general case this can lead to an additional Jacobian factor and nontrivial vielbein coefficients \( \xi_i^a \), resulting in anisotropic Fermi velocity.

%frag:  2012-07-12

% --------------------------------------------------------------------
\section{Coefficients  \( \Gamma_{iI} \)}\label{sec:Gammas}
% --------------------------------------------------------------------
In this section we discuss parameters \( \Gamma_{iI} \), specifying the Fermi point condition \eqref{cond-coord}. We shall establish the range of appropriate coefficients \( \Gamma_{iI} \), leading to point-like solutions for \eqref{cond-coord}. The appropriate parameters \( \Gamma_{iI} \) and \( m_{I} \) can take values within a range, which can be called \emph{moduli space}. Each connected component of the moduli space corresponds to an equivalence class of physical models which can be deformed into each other by continuous modifications of the elements of the adjacency matrix.

The equation \eqref{cond-coord} is homogeneous, therefore the consistent $\Gamma_{iI}$ and $m_I$ are determined up to a scale factor. Therefore it is convenient to fix the value of one non-zero parameter, say \( m_{I} \), for every \( I \). 

For a given \( I \),  one can represent graphically the equation \eqref{cond-coord} as a $(D+1)$-gon on the complex plane, consisting of $D$ sides of lengths $|\Gamma_{iI}|$ and orientation \( \e^{\ii K_{i}} \) as well as one horizontal side of length \(| m_I| \) (see Fig.\ref{fig:Dgons_all}a for an example). So, in the polygon representation, the upper case index \( I=1,\dots, D'/2 \) enumerates polygons, while the lower case latin $i=1,\dots, D$ sides inside one polygon. 

The appropriate coefficients \( \Gamma_{iI} \) and \( m_{I} \) are those for which all \( D'/2 \) polygons close by unique, up to discrete transformation, combination of angles \( K_{i} \). A continuous freedom in changing the angles \( K_i^* \), would mean that the respective Fermi variety is not a point but a higher dimensional structure.

% --------------------------------------------------------------------
%
\input{Dgons_all.TpX}
% --------------------------------------------------------------------

It is useful to point to a mechanical analogy  in terms of arm-and-hinge mechanisms, following from the polygon representation of equation \eqref{cond-coord}. If we normalize the polygons by rescaling their size by \( m_{I}^{-1} \), such that they will have a common base, we can build a mechanism made of arms of lengths \( \Gamma_{iI} \) connected within a polygon  by hinges allowing free motion within the two dimensional plane. The arms of different polygons having the some number \( i \) are kept strictly parallel because of common orientation \( K_{i} \), see Fig. \ref{fig:Dgons_all}b. The ``correct'' mechanism is one which is rigid, i.e. one in which no angle \( K_{i} \) can be modified without breaking at least one arm. 

Let us evaluate the minimal number of independent polygons which are necessary to fix the continuous freedom of \( K_{i} \). Any polygon, consisting of \( (D+1) \) edges of fixed lengths  can be deformed in the two-dimensional plane by \( (D-2) \) independent ways using its vertices as joints. Therefore, one polygon condition fixes two out of \( D \) angles \( K_{i} \). To fix all \( D \) values uniquely, we need \( D/2 \) independent polygon constraints if dimension is even, or \( (D+1)/2 \) ones if the dimension is odd, i.e. the number is \( D'/2 \). It is interesting to  point that this matches perfectly the dimension of the minimal Clifford algebra. This match appears both miraculous and beautiful since there is no known direct relation between the moduli of complex plane polygons and the Clifford algebras. 

In the following subsections we consider first solutions for two simplest, but most interesting cases: \( D=2 \) and \( 3 \), then we turn to the case of an arbitrary dimension.
 
% --------------------------------------------------------------------
% --------------------------------------------------------------------
\subsection{The case: $D=2$}
% --------------------------------------------------------------------
For $D=2$, a single equation \eqref{cond-coord}  fixes the Fermi point,
\begin{equation}\label{d2-Gamma}
 	\sum_{i}\Gamma_{i1}\e^{\ii K_i}+m_{1}=0.
\end{equation}
The holomorphic Clifford algebra generator is \( \Sigma_{1}= \sigma_{+} \), where \( \sigma_{+} \) is the Pauli matrix \( \sigma_{+}=\frac{1}{2}	(\sigma_{1}+ \ii\sigma_{2})\).

Equation \eqref{d2-Gamma} implies that two vectors $\Gamma_{i1}\e^{\ii K_i}$  form, together with the \( m_{1} \)-site  in the real direction, a triangle on the complex plane, see Fig. \ref{fig:triangle}.
% --------------------------------------------------------------------
\input{triangle.tpx}
% --------------------------------------------------------------------
The lengths of the three sites determine the orientation angles $K_{i}$,  uniquely, up to a reflection. Thus the elementary triangle sine rule reads:
\begin{equation}
 	\frac{\Gamma_{11}}{|\sin K_2|}=\frac{\Gamma_{21}}{|\sin K_1|}=\frac{m_{1}}{|\sin(K_2-K_1)|}.
\end{equation}

Then the choice,
\begin{equation}\label{nu1D3}
	\Gamma_{11}=\frac{m_{1} \sin K^*_2}{\sin(K^*_1-K^*_2)},\quad
	\Gamma_{21}=\frac{m_{1} \sin K^*_1}{\sin(K^*_2-K^*_1)},
\end{equation}
results in a system with two Fermi points located at $ \pm\mathbf{K}^*=\pm (K_1^*,K_2^*)$.

%%frag: 2012-7-9

Let us turn to the low energy fluctuations. The the low energy fluctuations around the Fermi vacua $\pm K^*$ are described by the effective Hamiltonian,
\begin{equation}
 	H_{\mathrm{le}}=\int\dd^2k \left(
	\psi_{+}^\dag (\alpha_{+} \cdot k) \psi_{+}
	+\psi_{-}^\dag (\alpha_{-} \cdot k) \psi_{-}
	\right),
\end{equation}
where,
\begin{multline}
 	\alpha_{\pm} \cdot k=\pm m\frac{\sin K_1^*\sin K_2^*}{\sin (K_2^*-K_1^*)}(-k_1+k_2)\sigma_1\\
+
	\frac{m}{\sin (K_2^*-K_1^*)} \left[-(\cos K_1^* \sin K_2^*) k_1+
	(\cos K_2^*\sin K_1^*) k_2\right]\sigma_2.
\end{multline}

Introducing the index related to the Fermi point sign \( \pm \), the Hamiltonian is cast in the form,
\begin{equation}
 	H_{\mathrm{le}}=\int\dd^2k\,
	\psi^\dag \hat{\alpha} \cdot k\, \psi , 
\end{equation}
where
\begin{multline}\label{skew2D}
 	\hat{\alpha} \cdot k=  m\frac{\sin K_1^*\sin K_2^*}{\sin (K_2^*-K_1^*)} (-k_1+k_2)\sigma_1 \otimes \sigma_3\\
+
	\frac{m}{\sin (K_2^*-K_1^*)}\left[-(\cos K_1^* \sin K_2^*) k_1+
	(\cos K_2^*\sin K_1^*) k_2\right]\sigma_2 \otimes \I .
\end{multline}

The  `Cartesian momenta' \( q_{x} \) and \( q_{y} \) are given by,
\begin{subequations}\label{k2Cartesian}
\begin{align}
 	q_x&= 
	\frac{m}{\sin (K_2^*-K_1^*)}\left[-(\cos K_1^* \sin K_2^*) k_1+
	(\cos K_2^*\sin K_1^*) k_2\right],\\
	q_y&=m \frac{\sin K_1^*\sin K_2^*}{\sin (K_2^*-K_1^*)} (k_1-k_2).
\end{align}
\end{subequations}

Introducing Dirac matrices and going back to the real space action as discussed in Section \ref{sec:FermiPts} we obtain the Dirac fermion action \eqref{genuine-Dirac} in \( 2+1 \) dimensions.

In fact, this general solution falls in the universality class of graphene. Thus, in the case of graphene the Fermi point is located at \( \mathbf{K}^{*}= (2\pi/3,4\pi/3)\). This corresponds to the equilateral triangle: \( \Gamma_{11}= \Gamma_{21}=m_{1}\equiv t \), where \( t \) is the nearest neighbor transition amplitude (see e.g. \cite{2009RvMP...81..109C} and \cite{Sochichiu:2010ns} for a review).
Since in the case of graphene the transition amplitudes are obeying the same type of symmetry as the physical lattice, the induced spacial metric \eqref{ind-metric} is proportional to the real space metric,
\begin{equation}
 	g_{ij}=v_{\mathrm{F}}^{2} \delta_{ij},
\end{equation}
where \( v_{ \mathrm{F}}=3ta/2 \) is the Fermi velocity in graphene, and \( a \) is lattice spacing.

\begin{figure}[tbp]
\begin{center}
 \def\svgwidth{6cm}
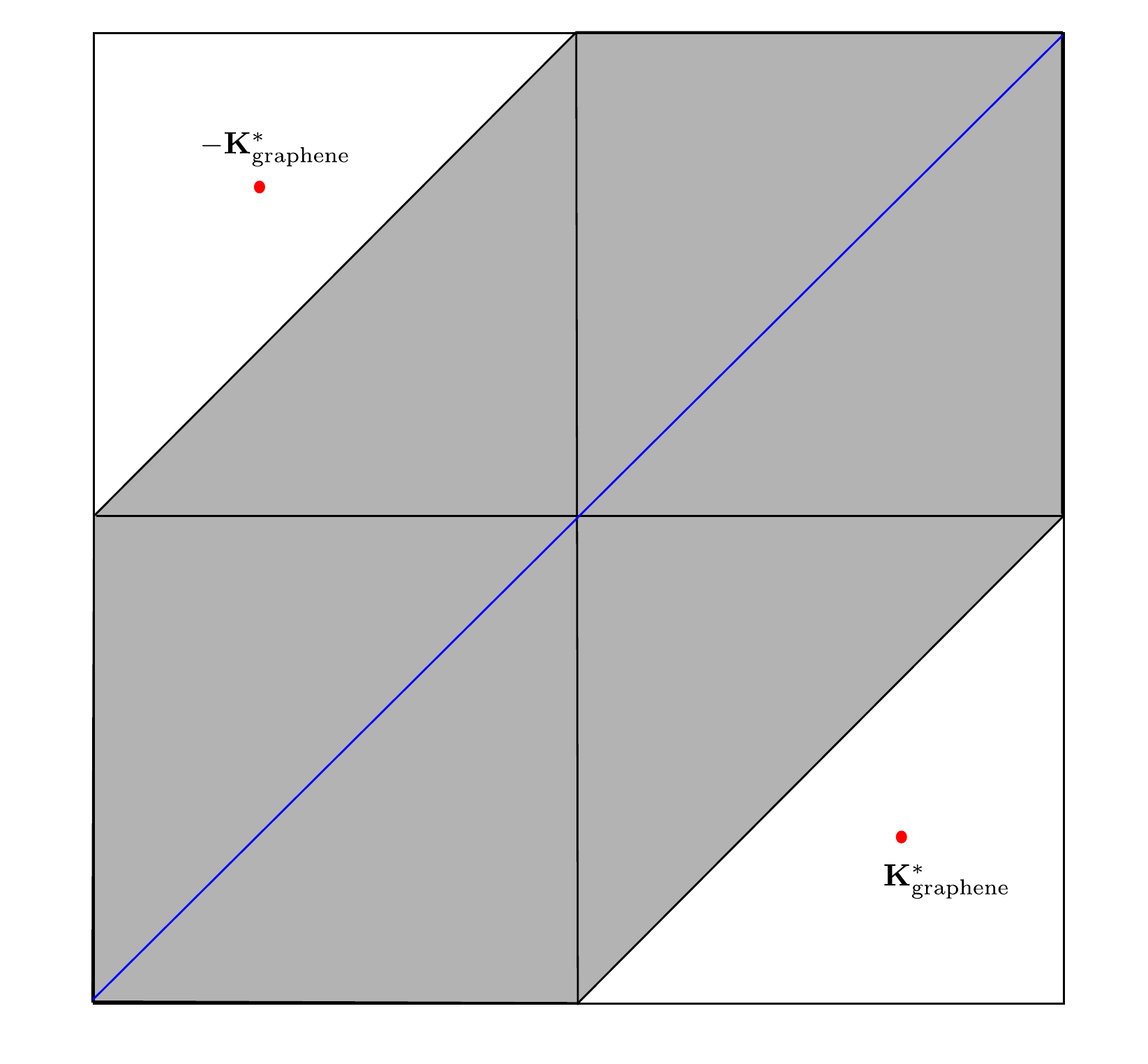
\end{center}
\caption{The two-dimensional Brillouin zone. The gray shaded area is the forbidden location for the Fermi point. 
Since the models with Fermi point locations \( \mathbf{K}^{*} \) and \( - \mathbf{K}^{*} \) are physically identical, only one triangular zone gives distinct graph model. The system becomes degenerate along the boundaries between allowed and forbidden zones as well as at the vertices. Due to periodicity the left side of the Brillouin zone should be glued to the right side and top to the bottom. Because of this the vertices of both triangles are, in fact, identified.}\label{Fig:2DBrillouin}
\end{figure}

The solution, we constructed in this section, has a set of singular points. The location of the Fermi points can not be arbitrary. By inspecting the possible orientations of the triangle edges in Fig.\ref{fig:triangle}, we come to the conclusion, that the momenta are constrained to the following regions,
\begin{equation}
	\begin{cases}
	0<K^{*}_{1}<\pi, \\
	-\pi< K^{*}_{2}<-\pi+K^{*}_{1},
	\end{cases}
	\quad\text{and}\quad
	\begin{cases}
	-\pi < K^{*}_{1}<0,\\
	\pi+K^{*}_{1}< K^{*}_{2}<\pi. 
	\end{cases}
\end{equation}
These two regions are connected at their corners, taking into account the periodicity of the momentum space, see the Fig.\ref{Fig:2DBrillouin}. In fact, since the models with Fermi point locations \(  \mathbf{K}^{*} \) and \( - \mathbf{K}^{*} \) are physically equivalent, independent solutions are coming from only one of the triangular zones.

% --------------------------------------------------------------------
% --------------------------------------------------------------------

The solution becomes singular as the Fermi point location approaches an edge or a vertex of the allowed zone.
Let us analyze these points in more details. 

A manifest singularity occurs along the edge for which \( \sin (K^{*}_{1}-K^{*}_{2})=0 \). If neither of \( \sin K^{*}_{i} \) vanish, the Fermi point is possible only in the case when \(m_{1}=0 \). In fact, this solution can be continuously approached from the regular one, given by \eqref{nu1D3}, by taking the limit
\begin{equation}\label{limit2D}
 	m_{1}\to t\sin(K^{*}_{1}-K^{*}_{2}),\qquad \sin (K^{*}_{1}-K^{*}_{2})\to 0.
\end{equation}
where \( t \) is the scaling parameter.

Applying the limit \eqref{limit2D} to the expression for the \( \hat{ \alpha} \) \eqref{skew2D}, we get,
\begin{equation}\label{singular2D1}
 	\hat{ \alpha}\cdot \mathbf{k}= t (k_{1}-k_{2})
	\left\{
	\sin^{2} K^{*} \sigma_{1} \otimes \sigma_{3}+\sin K^{*}\cos K^{*} \sigma_{2} \otimes \I
	\right\},
\end{equation}
where we have chosen the Fermi point to approach the location \( \mathbf{K}^{*}=(K^{*}, K^{*}\pm \pi ) \), with  \( K^{*}\neq 0,\pm\pi \). It is clear that the low energy limit of this case describes a one-dimensional infinitely degenerate Dirac fermion. The degeneracy of kinetic term is associated with respective degeneracy of induced metric.

Now, let us switch to the case when the Fermi point approaches one of the other two sides of the allowed zone. This corresponds to the limit \( \sin K^{*}_{i}\to 0 \) for one of \( i=1,2 \). The parameter \( m_{I} \) remains finite in this case, but the triangle in the Fig.\ref{fig:triangle} should flatten to the real axis, which means that the other component of the Fermi point momentum should also satisfy \( \Gamma_{j} \sin K^{*}_{j}=0 \). This means that unless the Fermi point is touching a vertex, the respective parameter \(  \Gamma_{i} \) should vanish. It is clear that vanishing of any parameter \( \Gamma_{i} \) means that any dependence on the momentum component \( k_{i} \) is automatically excluded in the low energy theory and the model is also one-dimensional.   

Now let us look what happens at the corners, i.e. when the sins of both Fermi momentum coordinates vanish. In this case one can observe that the conjugate Fermi points \( \mathbf{K}^{*}\leftrightarrow - \mathbf{K}^{*} \), merge into a single point. Therefore, the Fermi point degeneracy, which leads to doubling of components of the fermion disappears as well. The solution \eqref{nu1D3} is replaced in this case by,\footnote{Without loosing generality, we can take \( K^{*}_{1}=-K^{*}_{2}=\pi \), but allow \( \Gamma_{i} \) to be both positive and negative.}
\begin{equation}
 	\Gamma_{1}=\xi,\qquad \Gamma_{2}=m-\xi,
\end{equation}
where \( -\infty <\xi<+\infty \). The low energy fluctuations are described by,
\begin{equation}
 	\alpha\cdot \mathbf{k}=
	\left( \xi k_{1}+(m-\xi)k_{2}\right) \sigma_{2}.
\end{equation}
This corresponds to one dimensional Dirac fermion model as well.

Let us note however, that due to fusion of Fermi points in the case of the corner location, the protection mechanism due to conservation of the topological charge is lost. Therefore, the system is unstable with respect to small perturbations, and in the absence of a different mechanism of protection will be totally gapped in the case of interactions turned on (see \cite{VolovikLect.NotesPhys.718:31-732007}).

Let us conclude this subsection by pointing out that once the Fermi point location is inside of the allowed zone the model is not much different from the tight binding model for \emph{graphene} (see \cite{2009RvMP...81..109C} for a review), and falls into the same low energy equivalence class with it. In fact, the only physical Bravais lattice, which belongs to this solution is the hexagonal lattice. 
% --------------------------------------------------------------------

% --------------------------------------------------------------------
\subsection{$D=3$}\label{sec:3D}
% --------------------------------------------------------------------
In three dimensions, the polygon equations  \eqref{cond-coord} in general case are given by two quadrilaterals,
\begin{equation}\label{d3-Gamma}
 	\sum_{i=1}^{3} \Gamma_{iI}\e^{\ii K_i}+m_I=0, \qquad I=1,2.
\end{equation}

For non-zero \( m_{I}\), let us normalize the coefficients $\Gamma_{iI}$, by dividing them by $m_I$, such that they satisfy the normalized equation,
\begin{equation}
 	\sum_{i=1}^{3} \Gamma_{iI}\e^{\ii K_i}+1=0. 
\end{equation}
The original normalization can be restored by redefinition: $\Gamma_{iI}\to m_I \Gamma_{iI}$. 

Observe, that since the edges corresponding to the same direction have the same orientations on the complex plane, one can obtain both quadrilaterals from a single \emph{master triangle}, by cutting one angle of this triangle by parallel cuts, passing at different distances from the vertex, see the Fig.\ref{fig:quadrilateral}. This angle can be top or bottom, depending on angles \( K^{*}_{1} \) and \( K^{*}_{3} \). Then the problem of finding appropriate $\Gamma_{iI}$, reduces to solving the following equations:
% --------------------------------------------------------------------
\input{quadrilateral.tpx}
% --------------------------------------------------------------------
\begin{itemize}
\item[i.] Master triangle equation,
\begin{equation}\label{master-triangle}
	\Gamma_{1*}\e^{\ii K_1}+\Gamma_{3*}\e^{\ii K_3}+1=0.
\end{equation}
\item[ii.] Cut edge  equations,
\begin{equation}\label{cut-edge-triangle}
 	\eta_{1I}\e^{\ii K_1}+\eta_{2I}\e^{\ii K_2}+\eta_{3I}\e^{\ii K_3}=0.
\end{equation}
\end{itemize}
We formally wrote two cut edge equations \eqref{cut-edge-triangle}, but since they are homogeneous, only one of them is independent. As a result, the parameters $\eta_{iI}$ can differ by a scaling factor independent of space index, $\eta_{iI}=\lambda_I \eta_{i}$. Therefore, we can drop the capital index \( I \) in the cut edge equation \eqref{cut-edge-triangle}.

Once the master triangle and the cut edge equation are solved, the parameters \( \Gamma_{iI} \) can be found from \( \Gamma_{i*} \) and \( \eta_{iI}= \lambda_{I} \eta_{i} \) as follows,
\begin{equation}\label{master-cut-original}
 	\Gamma_{iI}=\Gamma_{i*}- \lambda_{I}\eta_{i},\quad i=1,3, \qquad \Gamma_{2I}= \lambda_{I}\eta_{2}.
\end{equation}
This solution comes in complete analogy with the two-dimensional case. Thus, for the master triangle equation we have the solution,
\begin{equation}
 	\Gamma_{1*}=\frac{\sin K^*_3}{\sin(K^*_1-K^*_3)},\qquad 
	\Gamma_{3*}=\frac{\sin K^*_1}{\sin(K^*_3-K^*_1)},
\end{equation}
while for the cut edge equation we can write down the following solution,
\begin{equation}
 	\eta_{1I}=\frac{ \lambda_{I} \sin(K^*_3-K^*_2)}{\sin(K^*_1-K^*_3)},\quad
	\eta_{2I}= \lambda_{I},\quad
	\eta_{3I}=\frac{ \lambda_{I} \sin(K^*_1-K^*_2)}{\sin(K^*_3-K^*_1)}.
\end{equation}
Combining all together, we have,
\begin{subequations}\label{gamma-sol}
\begin{align}
 	\Gamma_{1I}&=m_I \Gamma_{1*}-\eta_{1I}
	=\frac{m_I\sin K^*_3+ \lambda_I \sin(K^*_2-K^*_3)}{\sin(K^*_1-K^*_3)}
	,\\
	\Gamma_{2I}&=- \eta_{2I}=- \lambda_I ,\\
	\Gamma_{3I}&=m_I  \Gamma_{3*}-\eta_{3I}=
	\frac{-m_I\sin K^*_1+ \lambda_I \sin(K^*_1-K^*_2)}{\sin(K^*_1-K^*_3)}.
\end{align}
\end{subequations}%

Using the solution \eqref{gamma-sol}, we can write down the induced metric and vielbein coefficients \( \xi_{i}^{a} \) given by equations \eqref{ind-metric} and \eqref{embedd}, respectively. Due to the triangular origin of this solution its properties in the \( K_{1}-K_{3} \) plane mostly mimic that of two-dimensional solution.

Below we construct the solution based on elementary triangles in an arbitrary number of dimensions.

%% 2012-07-12 to scrap
%% 3D example

% --------------------------------------------------------------------
\subsection{Holomorphic solutions for arbitrary dimension.}
% --------------------------------------------------------------------
Inspired by the subtle relations of the polygon equations to Clifford algebras, let us find a solution valid for arbitrary dimension. As above, even and odd dimensions appear differently: The odd dimensional cases are essentially dimensional reductions from higher dimensional even cases. 

Consider the general polygon equation \eqref{cond-coord},
\begin{equation}\label{cond-coord2}
 	\sum_{i} \Gamma_{iI} \e^{\ii K_{i}}+m_{I}=0,\qquad I=1,\dots, D'/2,
\end{equation}
and the following Ansatz,
\begin{equation}
 	\Gamma_{iI}= \Gamma'_{I} \delta_{i,2I-1}+ \Gamma''_{I} \delta_{i,2I}.
\end{equation}
As a result of substitution, the polygon equations \eqref{cond-coord2}  split into \( D'/2 \) independent triangular equations,
\begin{equation}\label{cond-coord2triang}
 	\Gamma'_{I}\e^{\ii K_{2I-1}}+ \Gamma''_{I}\e^{\ii K_{2I}}+m_{I}=0.
\end{equation}
This Ansatz mimics the canonical form of the rotational matrix, which in an appropriate basis is a composition of elementary rotations of two-dimensional planes.

The general solution to equations \eqref{cond-coord2triang} we know already. It is given by equation \eqref{nu1D3}. In terms of notations of this section it spells,
\begin{equation}\label{sol-gen-triang}
 	\Gamma'_{I}= \frac{m_{I}\sin K^{*}_{2I}}{\sin(K^{*}_{2I-1}-K^{*}_{2I})},\qquad
	\Gamma''_{I}= \frac{m_{I}\sin K^{*}_{2I-1}}{\sin(K^{*}_{2I}-K^{*}_{2I-1})},
\end{equation}
where \( I=1,\dots,D'/2 \).

This solution leads to the following vielbein coefficients \eqref{embedd},
\begin{align}
 	\xi_{2I-1}^{2I-1}&= \frac{m_{I}\sin K^{*}_{2I-1}\sin K^{*}_{2I}}{\sin (K^{*}_{2I-1}-K^{*}_{2I})}, \qquad
	&\xi_{2I-1}^{2I}= \frac{m_{I}\sin K^{*}_{2I}\cos K^{*}_{2I-1}}{\sin (K^{*}_{2I-1}-K^{*}_{2I})}, \\
	\xi_{2I}^{2I-1}&= -\frac{m_{I}\sin K^{*}_{2I-1}\sin K^{*}_{2I}}{\sin (K^{*}_{2I-1}-K^{*}_{2I})}, \qquad
	&\xi_{2I}^{2I}= -\frac{m_{I}\sin K^{*}_{2I-1}\cos K^{*}_{2I}}{\sin (K^{*}_{2I-1}-K^{*}_{2I-1})},
\end{align} 
where \( I=1,\dots, D'/2 \). The low energy induced metric has  a diagonal \( 2 \times 2 \) block structure, with the block  \( g_{(I)} \) given by,
\begin{multline}
g_{(I)}= \frac{m_{I}^{2}}{\sin(K^{*}_{2I-1}-K^{*}_{2I})}\\
\times\begin{pmatrix}
\sin^{2} K^{*}_{2} & -\sin K^{*}_{1}\sin K^{*}_{2}\cos(K^{*}_{1}-K^{*}_{2})\\
-\sin K^{*}_{1}\sin K^{*}_{2}\cos(K^{*}_{1}-K^{*}_{2}) & \sin^{2} K^{*}_{1}
\end{pmatrix}.
\end{multline}

This solution depends explicitly on \( D'/2+D \) parameters \( m_{I} \) and \( K_{i} \).  
The three-dimensional solution from the Section \ref{sec:3D}, can be obtained from \eqref{sol-gen-triang}, by a redefinition of parameters and change of Clifford algebra basis. Although, we believe that any general configuration can be generated from the triangular solution \eqref{sol-gen-triang} by the change of the Clifford algebra basis or unitary rotation, we will present no proof here. 

% --------------------------------------------------------------------
\subsection*{Space diamond}
% --------------------------------------------------------------------
If we change the initial setup of the problem, removing the time and promoting the Hamiltonian \eqref{hamiltonian1} to the status of Euclidean action, we can reproduce Creutz space diamond fermions \cite{Creutz:2007af}  from the solution \eqref{sol-gen-triang}. Indeed, for \( D'=4 \), let us choose the solution with the symmetric ``graphene-like'' location of the Fermi points,
\begin{equation}
 	K^{*}_{1}=K^{*}_{3}=-K^{*}_{2}=-K^{*}_{4}=2\pi/3.
\end{equation}
This choice together with the choice for \( m_{I} \),
\begin{equation}
 	m_{1}=m_{2}\equiv t,
\end{equation}
yields the solution,
\begin{equation}
 	\Gamma_{I}=t.
\end{equation}
Using this solution as well as the representation of the \( D' \)-dimensional Clifford algebra \eqref{Cliff4D} constructed in the appendix, we get a model equivalent to the Creutz space diamond proposed in \cite{Creutz:2007af}.

The resulting Dirac fermion has an internal SU(\( 2 \)) global symmetry due to the fact that each Fermi point contributes a full Dirac Fermion rather than a Weyl fermion minimally allowed by the Nelsen--Ninomiya theorem  \cite{Nielsen:1980rz}.
The pure Dirac fermion can be obtained by an approach similar to Kogut--Sussking staggered fermions \cite{Kogut:1974ag}, which,  in the case of space diamond fermions, was successfully applied in \cite{Borici:2007kz}.
The idea is to leave only a Weyl fermion at a Fermi point. To do this one leaves only positive chirality part of the fermionic field \( a_{ \mathbf{n}} \) in the Hamiltonian \eqref{transl-inv} in odd lattice sites (\( \sum_{i}n_{i}= \)~odd), and only negative chirality part in even lattice sites (\(\sum_{i}n_{i}=\)~even). The chirality in this case should be defined with respect to the irreducible Clifford algebra constructed directly from \( \Sigma_{I} \). Each Fermi point in the low energy limit contributes a Weyl fermion of different chirality together resulting in a pure Dirac fermion. The same mechanism can be applied also in general context of our model.

% --------------------------------------------------------------------
\section{Conclusion}\label{sec:conclusion}
% --------------------------------------------------------------------
In this work we considered models of fermionic particles on graphs, which in the low energy limit flow to Dirac fermion model. A graph satisfying this properties is called Dirac lattice.  As a necessary condition for this is the appearance of linear Fermi points. If the low energy limit admits a rotational symmetry, this is also the sufficient condition. In our description the graphs are parametrized by the  adjacency matrix. Therefore, the Dirac lattice condition is expressed as a condition on adjacency matrix.

We show that the existence of a non degenerate linear Fermi point in \( D \) dimensional space implies that the energy spectrum of fluctuations around it belongs to the space of generators of the \( D \)-dimensional Clifford algebra. 

In a special Clifford algebra basis, which we construct, the condition of emergence of low energy Dirac fermion is translated into algebraic equations, which can be interpreted as polygon closure conditions on the complex plane. We provide a general solution to these polygon equations in the case of two and three dimensions. The last case is essentially the dimensional reduction of the four-dimensional case. We also construct a solution in arbitrary number of dimensions, which  is superposition of two-dimensional solutions. We call this solution \emph{holomorphic solution}, because it is built using the holomorphic basis in the momentum space.  We believe that this solution is also a general one.
 
It is interesting to point out that the constructions for even and odd dimensional spaces are considerably different. This is related to K-theory stability of the Fermi points in the two cases \cite{PhysRevLett.95.016405}. Here we extend the stability arguments of \cite{VolovikLect.NotesPhys.718:31-732007}, allowing us to obtain Dirac particles also in odd dimensional spaces, essentially as dimensional reductions of superior Clifford algebras. 

One of the objectives of this study was to develop the control of the properties of the emergent space-time and simple internal symmetries from the fermionic particle on a graph model. Next logical step will be to develop the construction of more complicated internal symmetries. In particular, it would be interesting to find the criteria for a graph to flow to a physically realistic gauge model like QCD or Standard model.

As a by-product, the developed technique can have some additional applications. One important application is lattice simulation. The developed approach can be used to construct lattice discretization for fermion containing models. Indeed, the constructed graph model is a discretization of Dirac fermion and inclusion of gauge or gravity interactions can be reached through the deformation of the transition amplitudes. It would be interesting to see, however, the computational efficiency of such discretization.

Another point worth understanding is the relation of our analysis with the results of the analysis for crystallographic lattices considered in \cite{Manes:2011jk}. The graph described by the adjacency matrix in our case does not necessarily satisfy any symmetry and are not restricted to correspond to a physical lattice. The physical form of the graph in our case is determined among others by the Clifford algebra representation. Finding which representation leads to a physical lattice and link them to the crystallographic groups is another interesting direction for a future research.
% --------------------------------------------------------------------
\subsection*{Acknowledgments}

I am grateful to Shinsuke Kawai for reading the first version of this manuscript and giving critical remarks. The revision of this paper was completed during my visits to KEK (Tsukuba), Swansea University and IFA (in Moldova). I wish to thank my hosts for the chance to speak about my work, useful comments and, of course, kind hospitality. I thank  Juan Luis Ma\~{n}es Palacios for very useful critical remarks and discussions.

This work was supported by the Korean NRF research project No.2010-0007637.

% --------------------------------------------------------------------
% --------------------------------------------------------------------

%\newpage

% ----------------------------------------------------------------
\appendix
% --------------------------------------------------------------------
\section{Construction of the Clifford algebra basis}\label{app:Cliff}
All possible Clifford algebra representations are well known \cite{Atiyah19643}. Here we show, that for some even dimensional space one can construct a holomorphic basis of Clifford algebra generators \( \Sigma_{I} \), \( I=1,\dots,D'/2 \) such that all matrices of this basis have real entries and satisfy the anti commutation relations \eqref{cliff-alg}.

Since we are paying special attention to the case of two and three spacial dimension, let us start with the cases of \( D'=2 \) and, respectively, \( D'=4 \).
% --------------------------------------------------------------------
\subsection*{ \( D'=2 \) case}
% --------------------------------------------------------------------
In the \( D'=2 \) case there is a single nilpotent matrix \( \Sigma_{1} \equiv \sigma_{+} \), where \( \sigma_{+} \), \( \sigma_{+}^{2}=0 \), is the standard su\( (2) \) spin raising operator,
\begin{equation}
 	\sigma_{+}=
	\begin{pmatrix}
 	0 & 1 \\
	0 & 0
	\end{pmatrix}.
\end{equation}
It gives rise to the familiar Pauli $\sigma$-matrices, which are generators of two-dimensional Clifford algebra:
\begin{equation}
 	\hat{\beta}_1 \equiv-\sigma_1=-(\sigma_{+}+\sigma_{-}),\qquad
	\hat{\beta}_2 \equiv-\sigma_2=\ii (\sigma_{+}+\sigma_{-}),
\end{equation}
where \( \sigma_{-}=\sigma_{+}^\dag \).

In the main text, due to Fermi point degeneracy, the Clifford algebra generators appear in a reducible representation,
\begin{equation}
 	 \beta_1=\hat{\beta}_1 \otimes \sigma_3, \qquad \beta_2=\hat{\beta}_1 \otimes\I.
\end{equation}
From these the matrix \( \gamma^0 \) is introduced according to Equation \eqref{gamma0},
\begin{equation}
 	\gamma^0=\beta_1 \beta_2=\ii \sigma_3 \otimes \sigma_3.
\end{equation}
Then, the Dirac gamma matrices are given by multiplication of the above \( \beta \)-matrices by \( \gamma^0 \),
\begin{equation}\label{gamma2d}
 	\gamma^1=-\gamma^0 \beta_1=\sigma_2 \otimes \I, \qquad 
	\gamma^2=-\gamma^0 \beta_2=-\sigma_1 \otimes \sigma_3.
\end{equation}
The general situation here is equivalent to the case of graphene (see an excellent review on graphene \cite{2009RvMP...81..109C}). The representation \eqref{gamma2d} is reducible. The decomposition of matrices \eqref{gamma2d} in terms of  a product of an irreducible matrix and internal symmetry generator can be found in \cite{Sochichiu:2010ns}. This situation corresponds to minimally doubled Dirac fermion.

% --------------------------------------------------------------------
\subsection*{ \( D'=4 \) case}
% --------------------------------------------------------------------
Let us turn to the case of two independent $\Sigma_I$ matrices, needed to describe the three- and four-dimensional cases. The algebra \eqref{sigma-matr} for \( D'=4 \) can be realized in terms of two fermionic operators acting on the fermionic oscillator basis $\ket{s_1,s_2}$,  according to the following rule,
\begin{subequations}\label{Cliff4D}
 \begin{align}
 	\Sigma_1^\dag\ket{0,0}&=\ket{1,0},\qquad &
	\Sigma_2^\dag\ket{0,0}&=\ket{0,1},\\
	\Sigma_1^\dag\ket{1,0}&=0,\qquad &
	\Sigma_2^\dag\ket{1,0}&=-\ket{1,1},\\
 	\Sigma_1^\dag\ket{0,1}&=\ket{1,1},\qquad &
	\Sigma_2^\dag\ket{0,1}&=0,\\
 	\Sigma_1^\dag\ket{1,1}&=0,\qquad &
	\Sigma_2^\dag\ket{1,1}&=0,
\end{align}

\end{subequations}
where $s_i=0,1$ are the occupation numbers.
Re-ordering the fermionic oscillator basis into the basis \( \mathbf{e}_i \), \( i=1,\dots,4 \) as follows,
\begin{equation}
 	\mathbf{e}_{2s_2+s_1+1}=\ket{s_1,s_2},
\end{equation}
we represent the matrices $\Sigma_I$ in the following form,
\begin{equation}
 	\Sigma_1=
	\begin{pmatrix}
 	0&1&0&0\\
	0&0&0&0\\
	0&0&0&1\\
 	0&0&0&0
	\end{pmatrix},\qquad
	\Sigma_2=
	\begin{pmatrix}
	0 & 0 & 1& 0\\
	0 & 0 & 0&-1\\
	0&0&0&0\\
 	0&0&0&0
	\end{pmatrix}.
\end{equation}

% --------------------------------------------------------------------
\subsection*{General \( D' \) case}
% --------------------------------------------------------------------
It is not difficult to generalize the above two constructions to the case of a general even dimension \( D' \). In order to do this let us consider the Hilbert space of \( D'/2 \) fermionic oscillators built from the `vacuum state' \( \ket{0} \), annihilated by 
\( \Sigma_I \),
\begin{equation}
 	\ket{\mathbf{s}}=\Sigma_{I_1}^\dag \Sigma_{I_2}^\dag\dots \Sigma_{I_s}^\dag\ket{0}, \qquad 	
	\Sigma_{I}\ket{0}=0,
\end{equation}
where the set of indices is ordered as follows, \( I_1<I_2<\dots<I_s \).
The matrix element for the operator $\Sigma_I$ in such a basis is given by,
\begin{equation}\label{SigmaD-matr}
 	\bra{\mathbf{s}'}\Sigma_I\ket{\mathbf{s}}=
	(-1)^{m_{\mathbf{s}'}}\delta_{\mathbf{s}'+\hat{I},\mathbf{s} },
\end{equation}
where \( m_{\mathbf{s}'} \) is the number of filled positions with \( I_k<I \), and \( \mathbf{s}'+\hat{I}  \) is the formal state which differs from \( \mathbf{s}' \) by an additional state at position $I$. If the state \( I \) is already filled in \( \mathbf{s}' \), the right hand side of \eqref{SigmaD-matr} vanishes automatically.

By construction, matrices \( \Sigma_I \) satisfy the algebra \eqref{sigma-matr} and have real-valued entries.

% --------------------------------------------------------------------
\bibliographystyle{unsrt}
\bibliography{nonfermi}
	
\end{document}